\documentclass[12pt,a4paper]{article}
\usepackage{a4wide}
\usepackage{amsmath}
\usepackage{amssymb}
\usepackage{epsfig}
\usepackage{subfigure}
\usepackage{exscale}
\usepackage{float}
\usepackage{bbm}
\usepackage[numbers,sort&compress]{natbib}
\usepackage{pst-plot, pstricks,pst-math}
\usepackage{fancybox,amssymb,color}
\usepackage{graphicx}
\usepackage{pstricks, color, graphicx, epsfig, psfrag}
\usepackage{amsfonts,amsmath,amssymb,slashed}
\usepackage{dsfont}
\usepackage{bbm,bm}
\usepackage{fancyhdr, a4wide}
\usepackage[english]{babel}
\usepackage{subfigure}

\makeatletter
\@addtoreset{equation}{section}
\makeatother

\title{Order Parameter and Magnetization of Antiferromagnets in Mutually Parallel Staggered and Magnetic Fields}

\author{Christoph P.\ Hofmann$^a$ \\ \\
\normalsize{$^a$ Facultad de Ciencias, Universidad de Colima} \\
\vspace{0.3cm}
\normalsize{Bernal D\'iaz del Castillo 340, Colima C.P.\ 28045, Mexico} \\}

\begin{document}
\maketitle

\begin{abstract} \normalsize

We explore the behavior of the order parameter and the magnetization of antiferromagnetic solids subjected to mutually parallel
staggered and magnetic fields. The effective field theory analysis of the partition function is taken up to the two-loop level, where the
magnon-magnon interaction comes into play. These interaction effects, however, are small. A phenomenon that comes rather unexpectedly is
that the finite-temperature magnetization increases with temperature when the strengths of the staggered and magnetic field are held
constant.

\end{abstract}

\maketitle

\section{Introduction}
\label{Intro}

The theoretical and experimental characterization of the thermal properties of antiferromagnetic solids has a very long history.
Theoretical investigations that include magnetic fields in the analysis of antiferromagnetic solids comprise, e.g., Refs.~\citep{Kub52,
Ogu60,ABK61a,Joe62,Fal64,ABP68,FP68,HH69,CS70a,Mor72,Gho73,LR74,LR75,AUW77,MM79,Fis89,BFD90,MG94,Pan98,Pan99,BS04,FS04,HSSK07,NS07,KSHK08,
NVS12,WHG14,NVSFS17}. However, the situation where the magnetic field is aligned with the staggered field -- the situation we consider in
this study -- has been explored rather scarcely. This motivates the present systematic investigation.

Our approach to antiferromagnetic solids is based on magnon effective field theory that offers a {\it systematic} analysis valid at low
temperatures. Regarding the evaluation of the partition function, we go up to the two-loop level, i.e., take into account effects induced
by the spin-wave interaction that start to come into play at this order of the effective expansion. Note that an analogous effective field
theory analysis dedicated to antiferromagnetic films -- rather than three-dimensional antiferromagnetic solids -- has been provided very
recently in Ref.~\citep{Hof19}.

It should be emphasized that in the present situation where magnetic and staggered fields are mutually parallel, the staggered field cannot
take arbitrarily small values. If the staggered field becomes too weak in comparison to the magnetic field, then the antiferromagnet
undergoes a spin-flop transition, i.e., it realizes another ground state configuration where the order parameter is oriented perpendicular
to the magnetic field. This other situation has been studied previously within effective field theory for three-dimensional
antiferromagnets as well as for antiferromagnetic films in Refs.~\citep{BH17,BH19} and Refs.~\citep{Hof17,Hof18}, respectively.

As it turns out, two-loop corrections to the partition function are small, such that the thermodynamics of antiferromagnetic solids in
external fields is described quite accurately by the non-interacting magnon gas. When magnetic and staggered fields are held constant, the
order parameter drops when temperature increases. While this behavior of the order parameter is expected on account of the thermal
fluctuations that are stronger at more elevated temperatures, the behavior of the magnetization comes quite as a surprise: the
magnetization grows when temperature increases when magnetic and staggered field strengths are held fixed.

The article is organized as follows. In Sec.~\ref{eft} we discuss antiferromagnetic solids subjected to external fields -- both from a
microscopic and an effective point of view. In Sec.~\ref{FreeEnergyDensity} we evaluate the two-loop free energy density and discuss the
relevant scales involved. In Sec.~\ref{stagMag} and Sec.~\ref{mag}, respectively, we study the low-temperature behavior of the order
parameter and the magnetization of antiferromagnetic solids exposed to mutually parallel magnetic and staggered fields. In
Sec.~\ref{conclusions} we then conclude. In an appendix we consider technical details regarding the derivation of the two-loop free energy
density.

\section{Effective versus Microscopic Point of View}
\label{eft}

Within the microscopic perspective, antiferromagnetic solids are described by the Hamiltonian
\begin{equation}
\label{HeisenbergZeemanH}
{\cal H} \, = \, - J \, \sum_{n.n.} {\vec S}_m \! \cdot {\vec S}_n \, - \, \sum_n {\vec S}_n \cdot {\vec H} \, - \, \sum_n (-1)^n {\vec S}_n
\! \cdot {\vec H_s} \, , \qquad \qquad J < 0 \, .
\end{equation}
While the first contribution represents the isotropic quantum Heisenberg model, the additional terms contain an external magnetic
(${\vec H}$) and staggered (${\vec H_s}$) field, respectively. For simplicity, we assume that the underlying lattice is bipartite, and
we restrict ourselves to nearest-neighbor interactions.

It should be pointed out that the effective field theory cannot be "derived" from the microscopic model. Rather, in order to construct the
effective Lagrangian for an antiferromagnetic solid, one has to identify the symmetries that are present in the underlying microscopic
model, and one has to identify the relevant low-energy degrees of freedom. The basic observation is that the isotropic Heisenberg
Hamiltonian is invariant under O(3) spin rotations, whereas this symmetry is spontaneously broken by the ground state configuration that is
only O(2)-invariant. Goldstone's theorem then predicts two independent low-energy excitations: the magnons or spin waves that one
identifies as the relevant degrees of freedom in the low-energy effective field theory.

In this article, however, we do not aim at a systematic introduction into effective field theory, but only touch upon a few essential
features necessary to comprehend the present calculation.\footnote{More detailed outlines on effective field theory, specifically for
antiferromagnets subjected to magnetic and staggered fields, can be found, e.g., in sections IX-XI of Ref.~\citep{Hof99a}. From a
conceptual point of view the articles \citep{Leu94a,ABHV14} may also be of interest.} In the effective description of antiferromagnets, the
two magnon fields $U^1$ and $U^2$ are two components of the unit vector $U^i$,
\begin{equation}
U^i = (U^0, U^a) \, , \quad U^0 = \sqrt{1 - U^a U^a} \, ,\qquad a = 1,2 \, , \quad i = 0,1,2 \, .
\end{equation}
The ground state corresponds to the configuration ${\vec U}_0 = (1,0,0)$. The low-energy excitations -- the magnons -- correspond to
fluctuations of the unit vector $\vec U$ around ${\vec U}_0$.

The effective field theory formalism is restricted to the low-energy domain: we are dealing with an expansion in powers of energy, momentum
and temperature. The effective Lagrangian hence consists of a derivative expansion, where the leading (order $p^2$) contribution --
${\cal L}^2_{eff}$ -- contains two (covariant) space-time derivatives,
\begin{equation}
\label{Leff2}
{\cal L}^2_{eff} = \mbox{$ \frac{1}{2}$} \rho_s D_{\mu} U^i D^{\mu} U^i + M_s H^i_s U^i \, ,
\end{equation}
with
\begin{equation}
D_0 U^i = {\partial}_0 U^i + {\varepsilon}_{ijk} H^j U^k \, , \qquad D_r U^i = {\partial}_r U^i \, , \qquad (r=1,2,3) \, .
\end{equation}
Note that the external magnetic field $H^i$ is incorporated in the time covariant derivative $D_0 U^i$. The staggered field $H^i_s$, on the
other hand, couples to the effective constant $M_s$ that is identified as the zero-temperature staggered magnetization. The second
effective constant -- $\rho_s$ -- is the spin stiffness.

The subsequent contribution in the effective Lagrangian (order $p^4$) is given by
\begin{eqnarray}
\label{Leff4}
{\cal L}^4_{eff} & = & e_1 (D_{\mu} U^i D^{\mu} U^i)^2 + e_2 (D_{\mu} U^i D^{\nu} U^i)^2
+ k_1 \frac{M_s}{\rho_s} (H_s^i U^i) (D_{\mu} U^k D^{\mu} U^k) \nonumber \\
& & + k_2 \frac{M_s^2}{\rho_s^2} (H_s^i U^i)^2 + k_3 \frac{M_s^2}{\rho_s^2} H_s^i H_s^i \, .
\end{eqnarray}
At this order of the effective expansion we have five next-to-leading order (NLO) effective constants. Unlike $M_s$ and $\rho_s$ they have
no direct physical interpretation. Of course, to make predictions on the basis of the effective field theory, one has to determine -- or at
least estimate -- the numerical values of $e_1, e_2, k_1, k_2, k_3$ (see below).

In this study we are interested in the case where magnetic and staggered fields are mutually aligned,
\begin{equation}
\label{externalFields}
{\vec H}_{||} = (H,0,0) \, , \qquad {\vec H}_s = (H_s,0,0) \, , \qquad H, H_s > 0 \, .
\end{equation}
In order to derive the dispersion relation for the antiferromagnetic spin waves in presence of these external fields, it is convenient to
first define two alternative independent magnon fields $u(x)$ and $u^{*}(x)$ as
\begin{equation}
\label{physicalMagnons}
u = U^1 + i U^2  \, , \qquad u^{*} = U^1 - i U^2 \, .
\end{equation}
The leading-order effective Lagrangian ${\cal L}^2_{eff}$ then gives rise to the magnon dispersion relations
\begin{eqnarray}
\label{disprelAFHparallel}
\omega_{+} & = & \sqrt{{\vec k \,}^2 + \frac{M_s H_s}{\rho_s}} + H \, , \nonumber \\
\omega_{-} & = & \sqrt{{\vec k \,}^2 + \frac{M_s H_s}{\rho_s}} - H \, ,
\end{eqnarray}
that are well-known in the condensed matter literature \citep{ABK61a,Nol86}. Note that one can define a "magnon mass" $M$,
\begin{equation}
M^2 = \frac{M_s H_s}{\rho_s} \, ,
\end{equation}
that is tied to the staggered field $H_s$. In the isotropic case, i.e., in the absence of external fields, the spin-wave spectrum is
characterized by two degenerate spin-wave excitations that obey the dispersion law
\begin{equation}
\label{disprelAF}
\omega(\vec k) \, = \, |{\vec k}| \, , \qquad {\vec k} = (k_1,k_2,k_3) \, .
\end{equation}

It is important to point out that the lower spin-wave branch, $\omega_{-}$ in Eq.~(\ref{disprelAFHparallel}), becomes negative, unless the
condition
\begin{equation}
\label{stabilityCondition}
H_s > \frac{\rho_s}{M_s} \, H^2
\end{equation}
is fulfilled. Throughout the present study we assume that this stability criterion is satisfied.\footnote{If the staggered field becomes
too weak, then the magnetic field $\vec H$ forces the staggered magnetization vector to move into a configuration orthogonal to $\vec H$.
The effective analysis of antiferromagnets in mutually perpendicular staggered and magnetic fields has been outlined in
Refs.~\citep{Hof17,BH17,Hof18,BH19}.}

The basic objects needed to evaluate the partition function perturbatively are the thermal propagators for the antiferromagnetic magnons.
We first construct the Euclidean propagators at zero temperature. Given the dispersion relations Eq.~(\ref{disprelAFHparallel}), the
dimensionally regularized $T$=0 propagators are
\begin{eqnarray}
\label{regpropHd2}
\Delta^{\pm} (x) & = & \int \frac{\mbox{d} p_4}{2 \pi} \int \frac{{\mbox{d}}^{d_s} p}{{(2 \pi)}^{d_s}} \,
\frac{e^{i({\vec p} \, {\vec x} - p_4 x_4)}}{p_4^2 + {\vec p \,}^2 + M^2 \pm 2 i H p_4 - H^2} \nonumber \\
& = & {\int}_{\!\!\!0}^{\infty} \mbox{d} \rho \, \int \frac{\mbox{d} p_4}{2 \pi} \int \frac{{\mbox{d}}^{d_s} p}{{(2 \pi)}^{d_s}} \,
e^{i({\vec p} \, {\vec x} - p_4 x_4)} e^{-\rho (p_4^2 + {\vec p \,}^2 + M^2 \pm 2 i H p_4 - H^2)} \, ,
\end{eqnarray}
where $d_s$ is the spatial dimension and represents the regularization parameter. In particular, at the origin $x$=0, the zero-temperature
propagators take the form
\begin{eqnarray}
\Delta^{\pm} (0) & = & \frac{1}{2 \sqrt{\pi}} {\int}_{\!\!\!0}^{\infty} \mbox{d} \rho \, \rho^{-\frac{1}{2}}
\int \frac{{\mbox{d}}^{d_s} p}{{(2 \pi)}^{d_s}} \,e^{-\rho ({\vec p \,}^2 + M^2)} \nonumber \\
& = & \frac{M^{d_s-1}}{2^{d_s+1} \pi^{\frac{d_s}{2}+\frac{1}{2}}} \, \Gamma\Big(-\frac{d_s}{2}+\frac{1}{2}\Big) \, .
\end{eqnarray}
Note that these expressions, valid at $x$=0, no longer depend on the magnetic field: magnon $u$ and magnon $u^{*}$ are represented by the
same propagator ${\Delta}^{+}(0) = {\Delta}^{-}(0)$. In fact, it is identical with the (pseudo-)Lorentz-invariant propagator $\Delta(0)$,
\begin{eqnarray}
\label{regprop}
\Delta(0) & = & \int \frac{{\mbox{d}}^d p}{{(2 \pi)}^d} \, \frac{1}{M^2 + p^2}
= {\int}_{\!\!\!0}^{\infty} \mbox{d} \rho \, (4 \pi \rho)^{-d/2} e^{-\rho M^2} \nonumber \\
& = & \frac{M^{d-2}}{2^d \pi^{\frac{d}{2}}} \, \Gamma\Big(1-\frac{d}{2}\Big) \, .
\end{eqnarray}
Taking the physical limit $d_s \to 3$ ($d \to 4$) generates ultraviolet singularities on account of the Gamma function. We will see below
that these divergences can be absorbed into NLO effective constants.

Based on the propagators at $T$=0, one then constructs the thermal propagators $G^{\pm}(x)$ via\footnote{Elementary features of effective
field theory at non-zero temperature can be found, e.g., in Sec.~III of Ref.~\citep{Hof17}. A systematic account of the perturbative
evaluation of the partition function is given in chapters 2 and 3 of the standard textbook Ref.~\citep{KG06}.}
\begin{equation}
\label{ThermalPropagators}
G^{\pm}(x) = \sum_{n = - \infty}^{\infty} \Delta^{\pm}({\vec x}, x_4 + n \beta) \, , \qquad \beta = \frac{1}{T} \, .
\end{equation}
The dimensionally regularized expressions take the form
\begin{equation}
\label{ThermalPropagators2}
G^{\pm}(x) = \sum_{n = - \infty}^{\infty} \, {\int}_{\!\!\!0}^{\infty} \mbox{d} \rho \, \int \frac{\mbox{d} p_4}{2 \pi}
\int \frac{{\mbox{d}}^{d_s} p}{(2 \pi)^{d_s}} \, e^{-ip_4 (x_4 + n \beta) + i {\vec p} \, {\vec x}} e^{-\rho(p_4^2+{\vec p \,}^2 + M^2 \pm 2 i H p_4 - H^2)}
\, .
\end{equation}
It should be noted that, unlike at $T$=0, these expressions do depend on the magnetic field, and are thus different for magnon $u$ and
magnon $u^{*}$, respectively,
\begin{equation}
\label{ThermalPropagators3}
G^{\pm}(x) = \frac{1}{2 \sqrt{\pi}} \, \sum_{n = - \infty}^{\infty} \, {\int}_{\!\!\!0}^{\infty} \mbox{d} \rho \, \int
\frac{{\mbox{d}}^{d_s} p}{(2 \pi)^{d_s}} \rho^{-\frac{1}{2}} e^{-\rho({\vec p \,}^2 + M^2)} e^{i {\vec p} \, {\vec x}} e^{-\frac{{(x_4 + n \beta)}^2}{4 \rho}}
e^{\mp H (x_4+ n \beta)} \, .
\end{equation}
Defining the dimensionless parameters $h$ and $\tilde m$ as
\begin{equation}
\label{hm}
h = \frac{1}{2 \sqrt{\pi}} \frac{H}{T} \, , \qquad {\tilde m} = \frac{1}{2 \sqrt{\pi}} \frac{M}{T} =
\frac{1}{2 \sqrt{\pi}} \frac{\sqrt{M_s H_s}}{\sqrt{\rho_s} T} \, ,
\end{equation}
the thermal propagators at the origin $x$=0 amount to
\begin{equation}
\label{ThermalPropagators4}
G^{\pm}(0) = \frac{T^{d_s-1}}{4 \pi} \sum_{n = - \infty}^{\infty} \, {\int}_{\!\!\!0}^{\infty} \mbox{d} \rho \,
\rho^{-\frac{d_s}{2}-\frac{1}{2}} e^{-\rho {{\tilde m}^2}} e^{-\frac{\pi n^2}{\rho}} e^{\mp 2 \sqrt{\pi} h n} \, .
\end{equation}
Performing the sum analytically, one ends up with a Jacobi theta function $\theta_3(u,q)$,
\begin{equation}
\label{ThermalPropagatorsJacobi}
G^{\pm}(0) = \frac{T^{d_s-1}}{4 \pi} \, {\int}_{\!\!\!0}^{\infty} \mbox{d} \rho \, \rho^{-\frac{d_s}{2}} e^{-\rho {\tilde m}^2} \,
\theta_3\Big( \pm \sqrt{\pi} h \rho, e^{- \pi \rho}  \Big) e^{\rho h^2} \, ,
\end{equation}
where
\begin{equation}
\label{Jacobi3}
\theta_3(u,q) = 1 + 2 \sum_{n=1}^{\infty} q^{n^2} \cos(2 n u) \, .
\end{equation}
Note that the distinction between $G^{+}(0)$ and $G^{-}(0)$ is obsolete: the sign of the magnetic field exponential in the symmetric sum
over $n$, Eq.~(\ref{ThermalPropagators4}), is irrelevant. Likewise, the function $\theta_3(u,q)$ is even in $u=\sqrt{\pi} h \rho$. We hence
simplify our notation by writing
\begin{equation}
{\hat G(0)} = G^{+}(0) = G^{-}(0) \, .
\end{equation}
The thermal propagator ${\hat G(0)}$ also contains the $n$=0 (zero-temperature) contribution. It turns out to be convenient in the
renormalization process (see below) to isolate the mere thermal part in ${\hat G}(0)$. To that end we subtract the $n$=0 term, and define
the quantity ${\hat g}_1$ as
\begin{equation}
\label{ThermalPropagatorsg1}
{\hat g}_1 \equiv {\hat G}(0) - \Delta(0) \, .
\end{equation}
The kinematical Bose function ${\hat g}_1$,
\begin{equation}
\label{g1Bose}
{\hat g}_1 = \frac{T^{d_s-1}}{4 \pi} \, {\int}_{\!\!\!0}^{\infty} \mbox{d} \rho \, \rho^{-\frac{d_s}{2}-\frac{1}{2}} e^{-\rho {\tilde m}^2}
\Bigg\{ \sqrt{\rho} \, \theta_3\Big( \sqrt{\pi} h \rho, e^{- \pi \rho} \Big) e^{\rho h^2} - 1 \Bigg\} \, ,
\end{equation}
is well-defined in the physical limit $d_s \to 3$ and the numerical evaluation of
\begin{equation}
\label{g1BoseD3}
{\hat g}_1 = \frac{T^2}{4 \pi} \, {\int}_{\!\!\!0}^{\infty} \mbox{d} \rho \, \rho^{-2} e^{-\rho {\tilde m}^2}
\Bigg\{ \sqrt{\rho} \, \theta_3\Big( \sqrt{\pi} h \rho, e^{- \pi \rho} \Big) e^{\rho h^2} - 1 \Bigg\} \qquad \qquad (d_s = 3)
\end{equation}
poses no problems. Finally let us introduce the dimensionless Bose function ${\hat h}_1$ by
\begin{equation}
\label{defh1}
{\hat h}_1 = \frac{{\hat g}_1}{T^2} \, . 
\end{equation}

\section{Free Energy Density}
\label{FreeEnergyDensity}

\begin{figure}
\begin{center}
\includegraphics[width=15cm]{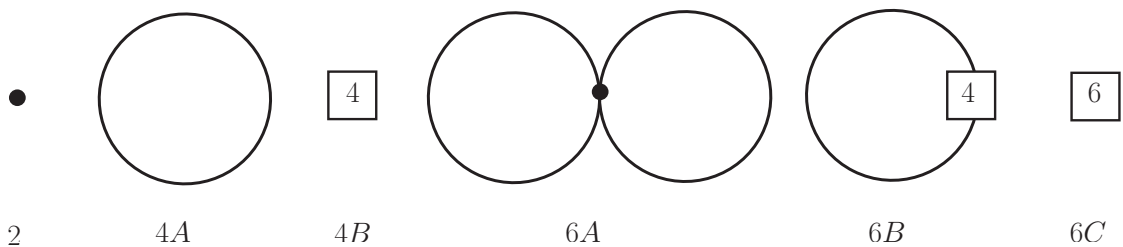}
\end{center}
\caption{Feynman graphs for the partition function describing antiferromagnetic solids in mutually parallel magnetic and staggered fields,
up to order $T^6$. Filled circles depict vertices from ${\cal L}^2_{eff}$. The numbers 4 and 6 refer to the subleading pieces
${\cal L}^4_{eff}$ and ${\cal L}^6_{eff}$ in the effective Lagrangian.}
\label{figure1}
\end{figure}

The thermal properties of antiferromagnetic solids in mutually parallel magnetic and staggered fields can be extracted from the partition
function (free energy density). The set of diagrams that we need to evaluate up to two-loop order is shown in Fig.~\ref{figure1}. The tree
graphs 2, 4B and 6C -- since they do not involve any thermal propagator lines -- just yield temperature-independent contributions. The
leading finite-temperature contribution in the free energy density originates from the one-loop diagram $4A$ (order $p^4 \propto T^4$).
Much like the next-to-leading order one-loop graph 6B (order $p^6 \propto T^6$), it describes non-interacting magnons. The effect of the
spin-wave interaction in the free energy density enters through the two-loop diagram 6A (order $p^6 \propto T^6$). The essential point is
that each additional magnon loop in a diagram suppresses the respective diagram by two powers of temperature, which puts the effective
low-temperature expansion on systematic grounds. An important observation is that the parallel magnetic field -- unlike an orthogonal
magnetic field (see Refs.~\citep{Hof17,BH17}) -- does not give rise to additional Feynman graphs (as compared to antiferromagnets in zero
magnetic field): ${\vec H}_{||}$ merely manifests itself in the thermal propagators.

The evaluation of the above Feynman diagrams is presented in Appendix \ref{appendixA} in detail. Here we just quote the final renormalized
representation for the two-loop free energy density:
\begin{eqnarray}
z & = & z_0 - {\hat g}_0 + \frac{H}{\rho_s} \, {\hat g}_1 \, \frac{\partial {\hat g}_0}{\partial H}
- \frac{H^2}{\rho_s}{( {\hat g}_1)}^2
- \frac{{\overline k}_2 - {\overline k}_1}{16 \pi^2} \frac{M^2_s H^2_s}{\rho^3_s} \, {\hat g}_1
- \frac{{\overline k}_1}{16 \pi^2} \, \frac{H M_s H_s}{\rho^2_s} \frac{\partial {\hat g}_0}{\partial H} \nonumber \\
& & + \frac{{\overline k}_1}{8 \pi^2} \, \frac{H^2 M_s H_s}{\rho^2_s} \, {\hat g}_1 + {\cal O}(p^8) \, .
\end{eqnarray}
The zero-temperature contribution $z_0$ reads
\begin{equation}
\label{z0}
z_0 = - M_s H_s + \frac{ {\overline k}_2 - 2{\overline k}_3}{32 \pi^2} \, \frac{M^2_s H^2_s}{\rho^2_s}
- \frac{M^2_s H^2_s}{64 \pi^2 \rho^2_s} + {\cal O}(p^6) \, .
\end{equation}
The quantities ${\overline k}_i \ (i=1,2,3)$ are renormalized NLO effective constants (see below). The thermal part of the free energy
density involves the kinematical function ${\hat g}_0$. It can be obtained from the Bose function ${\hat g}_1$ defined in
Eq.~(\ref{g1BoseD3}) through
\begin{equation}
\label{derg0}
{\hat g}_1 = - \frac{\mbox{d} {\hat g}_0}{\mbox{d} M^2} \, ,
\end{equation}
and thus reads
\begin{equation}
\label{g0Bose}
{\hat g}_0 = T^4 \, {\int}_{\!\!\!0}^{\infty} \mbox{d} \rho \, \rho^{-3} e^{-\rho {\tilde m}^2}
\Bigg\{ \sqrt{\rho} \, \theta_3\Big( \sqrt{\pi} h \rho, e^{- \pi \rho}  \Big) e^{\rho h^2} - 1 \Bigg\} \qquad \qquad (d_s = 3) \, .
\end{equation}
Remember that the parameters $h$ and ${\tilde m}$, defined in Eq.~(\ref{hm}), contain the magnetic and the staggered field. It is again
convenient to define the associated dimensionless Bose function ${\hat h}_0$ as
\begin{equation}
\label{defh0}
{\hat h}_0 = \frac{{\hat g}_0}{T^4} \, .
\end{equation}

The free energy density involves the three parameters $T, H, H_s$. By design, the low-energy effective field theory is restricted to the
domain where these parameters are small. The non-thermal scale with respect to which "small" has to be defined is fixed by the scale
present in the underlying microscopic model: the exchange integral $J$. Alternatively, "small" may be defined by the thermal scale $T_N$:
the N\'eel temperature. As discussed in Ref.~\citep{BH19}, $T_N$ can be estimated as
\begin{equation}
T_N \approx 3.5 \, \sqrt{\rho_s} \, .
\end{equation}
Furthermore, in case of the simple cubic spin-$\frac{1}{2}$ antiferromagnet, the spin stiffness in turn is related to the exchange integral
by (see Ref.~\citep{Hof99b})
\begin{equation}
\rho_s \approx 0.37 {|J|}^2 \, ,
\end{equation}
such that the two scales -- as expected -- are of the same order of magnitude,
\begin{equation}
T_N \approx 2.1 |J| \, .
\end{equation}
These considerations then translate into the statement that the domain of validity of our effective analysis can be described by three
dimensionless parameters,
\begin{equation}
\label{definitionRatios}
t \equiv \frac{T}{\sqrt{\rho_s}} \, , \qquad
m_H \equiv \frac{H}{\sqrt{\rho_s}} \qquad
m \equiv \frac{\sqrt{M_s H_s}}{\rho_s} \, ,
\end{equation}
that all ought to be small. In the present study, to be specific, we choose
\begin{equation}
\label{domain}
t, \, m_H, \, m \ \lesssim 0.6 \, .
\end{equation}

According to the preceding section, staggered and magnetic field strengths are subjected to the stability criterion
Eq.~(\ref{stabilityCondition}). In all plots that follow, we implement this criterion by restricting ourselves to the parameter region
\begin{equation}
m > m_H + \delta \, , \qquad  \delta = 0.1 \, .
\end{equation}
We are then on the safe side where our effective field theory representations are valid.

In $d_s$=3 -- and expressed in terms of $m, m_H$ and $t$ -- the kinematical functions ${\hat h}_0$ and ${\hat h}_1$, Eq.~(\ref{defh0}) and
Eq.~(\ref{defh1}), amount to
\begin{eqnarray}
{\hat h}_0 & = & {\int}_{\!\!\!0}^{\infty} \mbox{d} \rho \, \rho^{-3} e^{-\rho m^2/4 \pi t^2}
\Bigg\{ \sqrt{\rho} \, \theta_3\Big( \frac{m_H \rho}{2t}, e^{-\pi \rho} \Big) e^{\rho m_H^2/4 \pi t^2} - 1 \Bigg\} \, , \nonumber \\
{\hat h}_1 & = & \frac{1}{4 \pi} \, {\int}_{\!\!\!0}^{\infty} \mbox{d} \rho \, \rho^{-2} e^{-\rho m^2/4 \pi t^2}
\Bigg\{ \sqrt{\rho} \, \theta_3\Big( \frac{m_H \rho}{2t}, e^{-\pi \rho} \Big) e^{\rho m_H^2/4 \pi t^2} - 1 \Bigg\} \, .
\end{eqnarray}
The low-temperature expansion of free energy density then takes the form
\begin{eqnarray}
\label{freeEnergyDensity}
& & z = z_0 + {\hat z}_1 \, T^4 + {\hat z}_2 \, T^6 + {\cal O}(T^8) \, , \nonumber \\
& & \quad {\hat z}_1 = - {\hat h}_0 \, , \nonumber \\
& & \quad {\hat z}_2 = \Bigg[ m_H t^2 \, {\hat h}_1 \, \frac{\partial {\hat h}_0}{\partial m_H}
- m^2_H \, {({\hat h}_1 )}^2
- \frac{{\overline k}_2 - {\overline k}_1}{16 \pi^2} \, \frac{m^4}{t^2} \, {\hat h}_1
- \frac{{\overline k}_1}{16 \pi^2} \, m^2 m_H \frac{\partial {\hat h}_0}{\partial m_H} \nonumber \\
& & \qquad \qquad + \frac{{\overline k}_1}{8 \pi^2} \, \frac{m^2 m^2_H }{t^2}\, {\hat h}_1 \Bigg] \, \frac{1}{\rho_s t^2}\, .
\end{eqnarray}
The dominant contribution is of order $T^4$ (free magnon gas), while the next-to-leading contribution is of order $T^6$ and contains the
spin-wave interaction.

\begin{figure}
\begin{center}
\hbox{
\includegraphics[width=8.2cm]{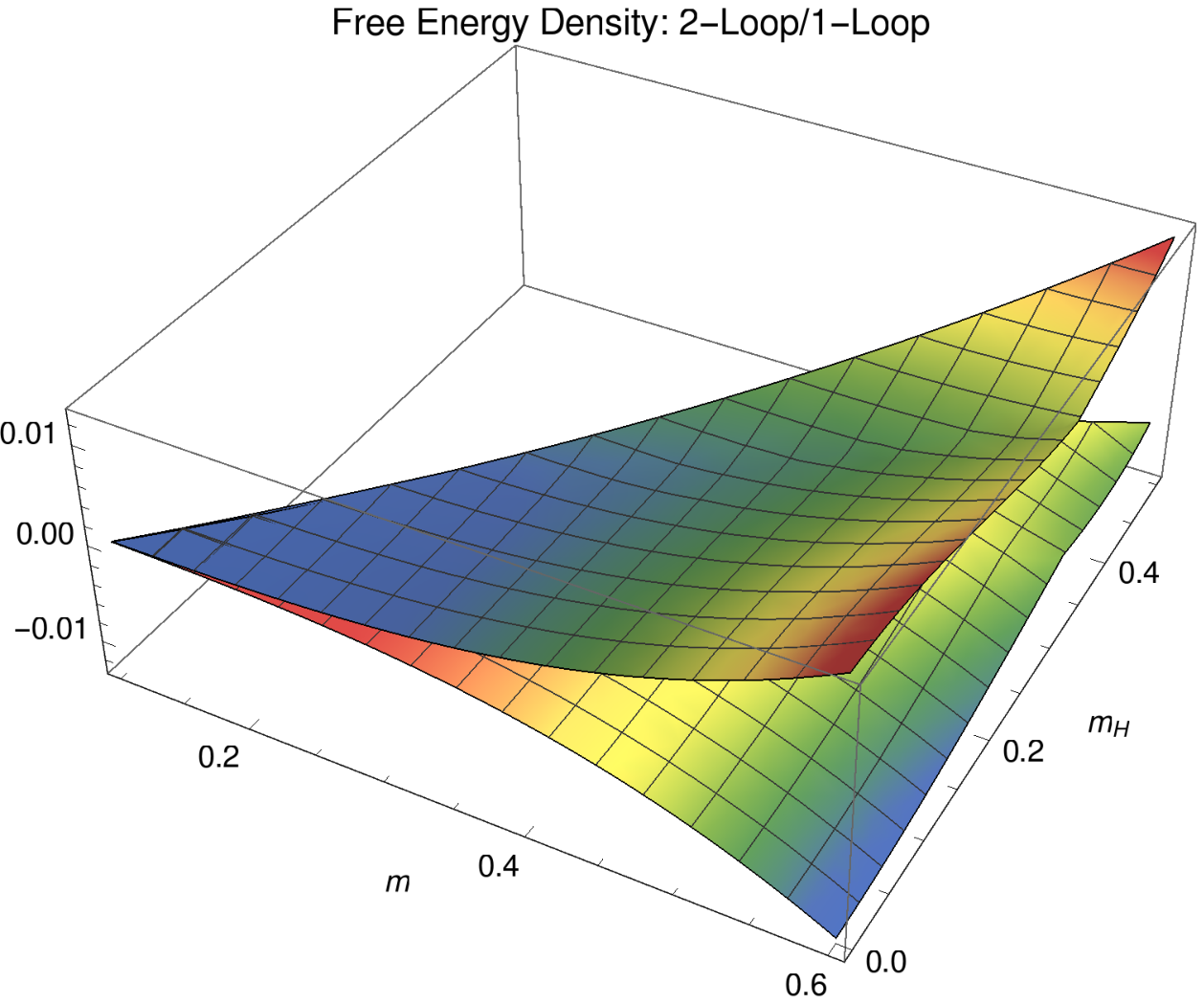} 
\includegraphics[width=8.2cm]{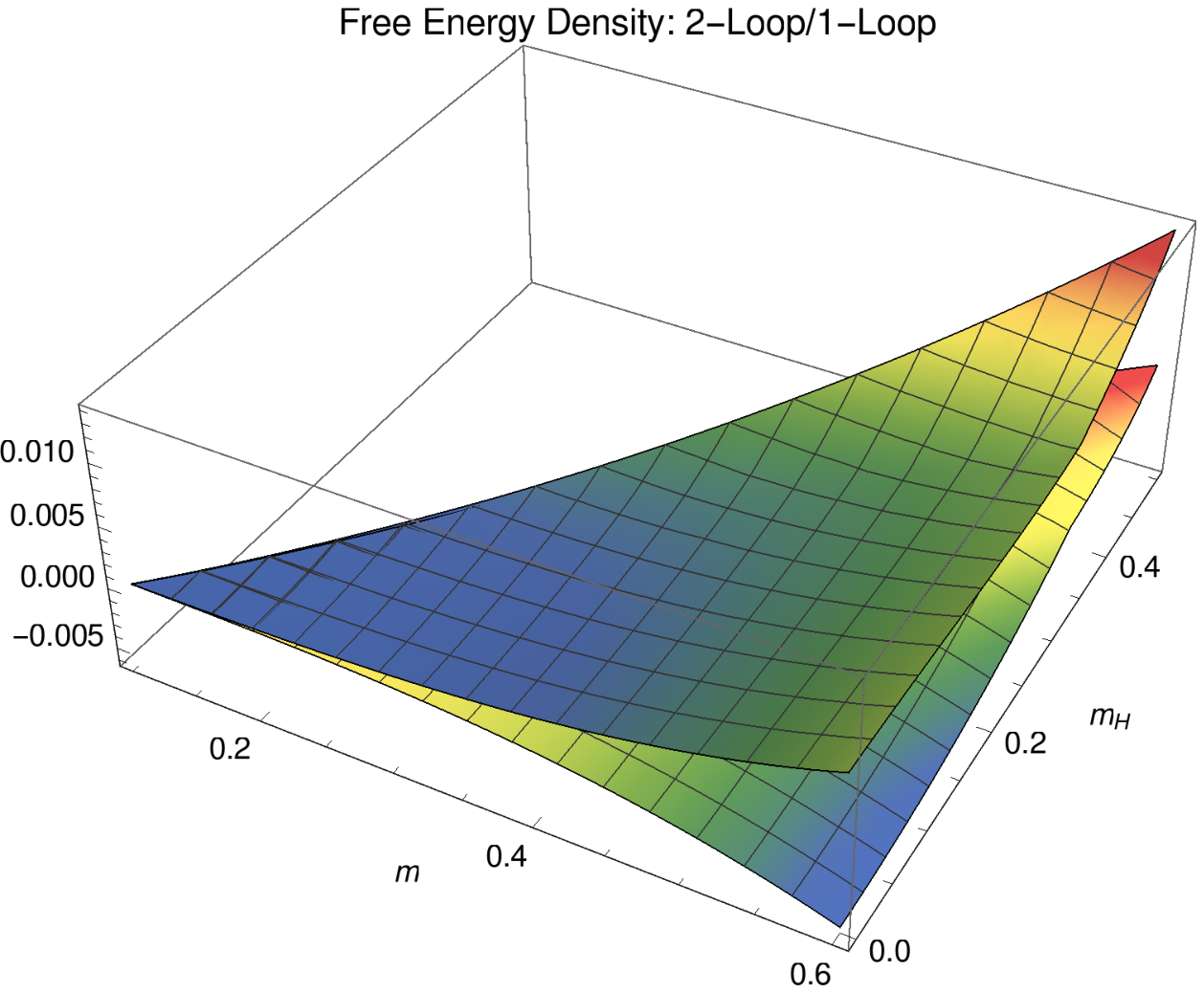}}
\end{center}
\caption{[Color online] Two-loop free energy density relative to one-loop free energy density, Eq.~(\ref{figFED2}), at the temperatures
$t = 0.3$ (left) and $t = 0.5$ (right): Dependence on the magnetic ($m_H$) and staggered ($m$) field. For both temperatures, maximal and
minimal two-loop contributions have been obtained by scanning ${\overline k}_1$ and ${\overline k}_2$ according to
Eq.~(\ref{scanningk1k2}).}
\label{figure2}
\end{figure}

In order to assess the magnitude of the two-loop correction ${\hat z}_2 \, T^6$, we should know the numerical values of the NLO effective
constants ${\overline k}_1$ and ${\overline k}_2$ that appear in the coefficient ${\hat z}_2$. These could in principle be determined by
matching Monte Carlo simulations or microscopic calculations with our effective results. Unfortunately, these options are not available to
the best of our knowledge, such that we have to base our numerical analysis on estimates for ${\overline k}_1$ and ${\overline k}_2$. While
the sign of these constants remains open, we can estimate their magnitude on general effective field theory arguments. According to
Ref.~\citep{BH17}, they are of "natural" size, i.e.,\footnote{Note that the quantities ${\overline k}_i$ are of natural size at the fixed
renormalization scale $\mu \approx 1.61 \sqrt{\rho_s}$ according to Ref.~\citep{Hof17b}. They depend logarithmically on the parameter $m$
as ${\overline k}_i = - \log{(\frac{m}{1.61})}^2 + {\hat k}_i$, where the quantities ${\hat k}_i$ are numbers of order one.}
\begin{equation}
{\overline k}_1, {\overline k}_2 \approx 1 \, .
\end{equation}
We then proceed by scanning ${\overline k}_1$ and ${\overline k}_2$ in the interval
\begin{equation}
\label{scanningk1k2}
\{ {\overline k}_1, {\overline k}_2 \} \ \subset \ [ -5, 5 ] \, ,
\end{equation}
and obtain a set of surfaces for the free energy density $z(t,m,m_H)$. Out of these scans we then choose the two extreme situations, namely
the minimal and maximal two-loop corrections for each point parameterized by $t,m,m_H$. This gives us estimates of the lower and upper
bounds for the two-loop contribution. In Fig.~\ref{figure2}, we depict the dimensionless ratio
\begin{equation}
\label{figFED2}
\frac{{\hat z}_2 T^2}{|{\hat z}_1|}
\end{equation}
for the two  temperatures $T/\sqrt{\rho_s} = 0.3$ (left) and $T/\sqrt{\rho_s} = 0.5$ (right). In either case the order-$T^6$ corrections are
at most a few percent as compared to the dominant free magnon gas contribution ${\hat z}_1$.

\section{Order Parameter}
\label{stagMag}

The order parameter -- the staggered magnetization -- is defined as
\begin{equation}
M_s(T,H_s,H) = - \frac{\partial z(T,H_s,H)}{\partial H_s} \, ,
\end{equation}
where $z(T,H_s,H)$ is the free energy density. The low-temperature series involves even powers of the temperature
\begin{equation}
\label{OPAF}
M_s(T,H_s,H) = M_s(0,H_s,H) + {\tilde \sigma}_1 T^2 + {\tilde \sigma}_2 T^4 + {\cal O}(T^6) \, ,
\end{equation}
with coefficients
\begin{eqnarray}
{\tilde \sigma}_1(T,H_s,H) & = & - \frac{M_s}{\rho_s} \, {\hat h}_1 \, , \nonumber \\
{\tilde \sigma}_2(T,H_s,H) & = & \frac{M_s}{\rho_s^2} \, \Bigg\{ m_H {\hat h}_2 \, \frac{\partial{\hat h}_0}{\partial m_H}
+ m_H {\hat h}_1 \, \frac{\partial{\hat h}_1}{\partial m_H}
- \frac{2 m_H^2}{t^2} \, {\hat h}_1 {\hat h}_2
+ \frac{{\overline k}_2-{\overline k}_1}{8 \pi^2} \, \frac{m^2}{t^2} \, {\hat h}_1 \nonumber \\
& & - \frac{{\overline k}_2-{\overline k}_1}{16 \pi^2} \, \frac{m^4}{t^4} \, {\hat h}_2
+ \frac{{\overline k}_1}{16 \pi^2} \, m_H \frac{\partial {\hat h}_0}{\partial m_H}
- \frac{{\overline k}_1}{16 \pi^2} \, \frac{m^2 m_H}{t^2} \, \frac{\partial{\hat h}_1}{\partial m_H} \nonumber \\
& & - \frac{{\overline k}_1}{8 \pi^2} \, \frac{m^2_H}{t^2} {\hat h}_1
+ \frac{{\overline k}_1}{8 \pi^2} \, \frac{m^2 m^2_H}{t^4} {\hat h}_2
- \frac{m_H}{16 \pi^2} \, \frac{\partial {\hat h}_0}{\partial m_H}
+ \frac{m^2_H}{8 \pi^2 t^2 } \, {\hat h}_1 \Bigg\} \, .
\end{eqnarray}
The kinematical function ${\hat h}_2$,
\begin{equation}
{\hat h}_2 = \frac{1}{16 \pi^2} \, {\int}_{\!\!\!0}^{\infty} \mbox{d} \rho \, \rho^{-1} e^{-\rho m^2/4 \pi t^2}
\Bigg\{ \sqrt{\rho} \, \theta_3\Big( \frac{m_H \rho}{2t}, e^{- \pi \rho} \Big) e^{\rho m^2_H/4 \pi t^2} - 1 \Bigg\} \, ,
\end{equation}
is dimensionless, and connected to ${\hat g}_1$ as
\begin{equation}
{\hat h}_2 = {\hat g}_2 = - \frac{\mbox{d} {\hat g}_1}{\mbox{d} M^2} \, .
\end{equation}

We first consider the order parameter at zero temperature,
\begin{eqnarray}
\label{OPzeroT}
M_s(0,H_s,H) & = & - \frac{\partial z(0,H_s,H)}{\partial H_s} = - \frac{\partial z_0}{\partial H_s} \nonumber \\
& = & M_s \Bigg\{ 1 - \frac{{\overline k}_2 - 2 {\overline k}_3}{16 \pi^2 \rho^2_s} \, M_s H_s  + {\cal O}(p^6) \Bigg\} \, .
\end{eqnarray}
The quantity $M_s = M_s(0,0,0)$ is the $T$=0 order parameter without external fields present. Unlike in the case of antiferromagnetic films
($d_s$=2) where the NLO effective constants only appear at NNLO in the zero-temperature order parameter (see Ref.~\citep{Hof19}), here they
already emerge at next-to-leading order. Whereas in $d_s$=2 these NLO constants have been determined by loop-cluster algorithms (for the
square lattice), here in $d_s$=3 their numerical values are not available and, again, we have to stick to our estimates. Still, knowing
that these are of order unity, the NLO correction in Eq.~(\ref{OPzeroT}) is small. Note that the magnetic field does not yet show up at
NLO: it starts manifesting itself in the NNLO (order $p^6$) corrections that also contain NNLO effective constants from ${\cal L}^6_{eff}$
(tree level diagram 6C).

\begin{figure}
\begin{center}
\hbox{
\includegraphics[width=8.2cm]{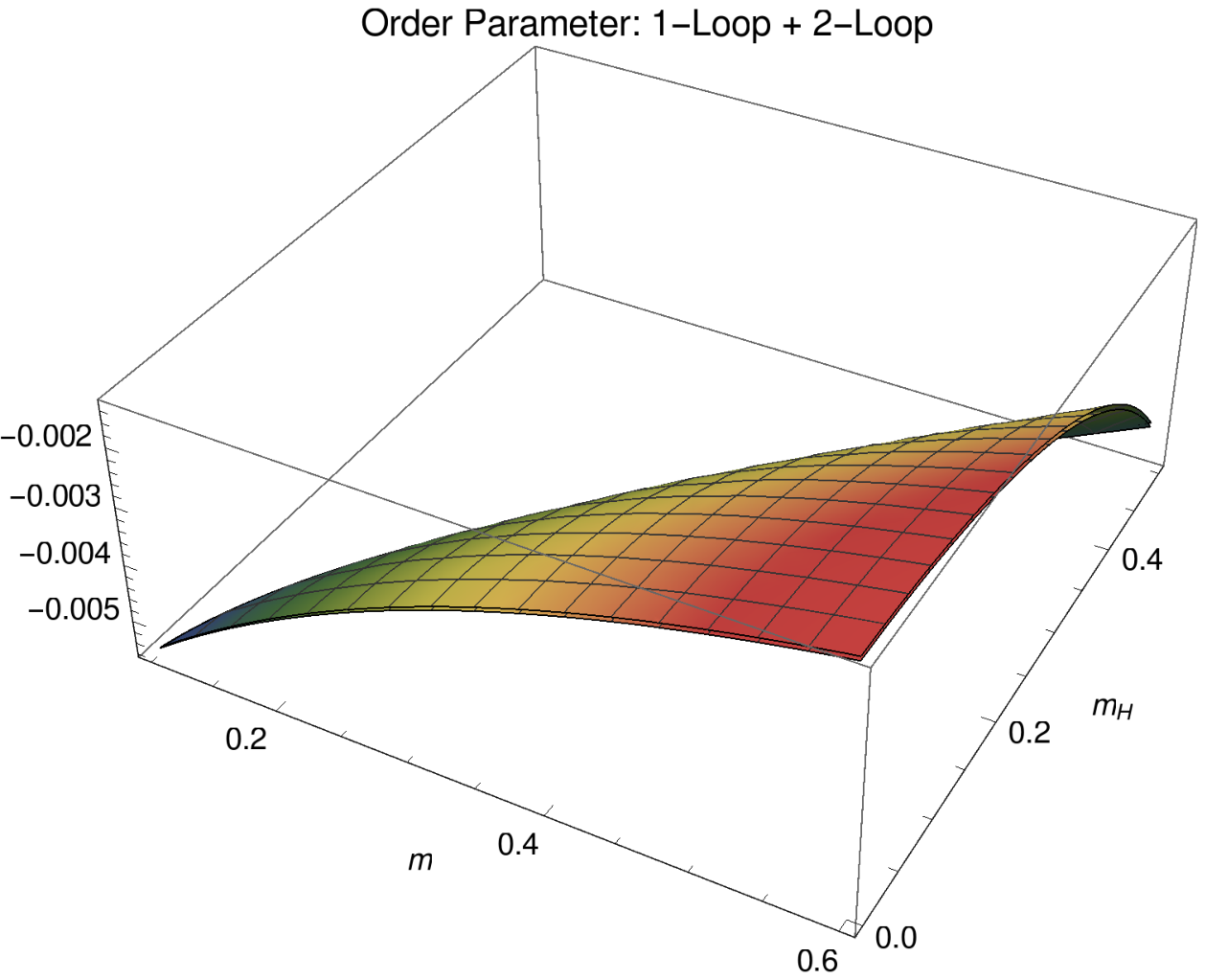} 
\includegraphics[width=8.2cm]{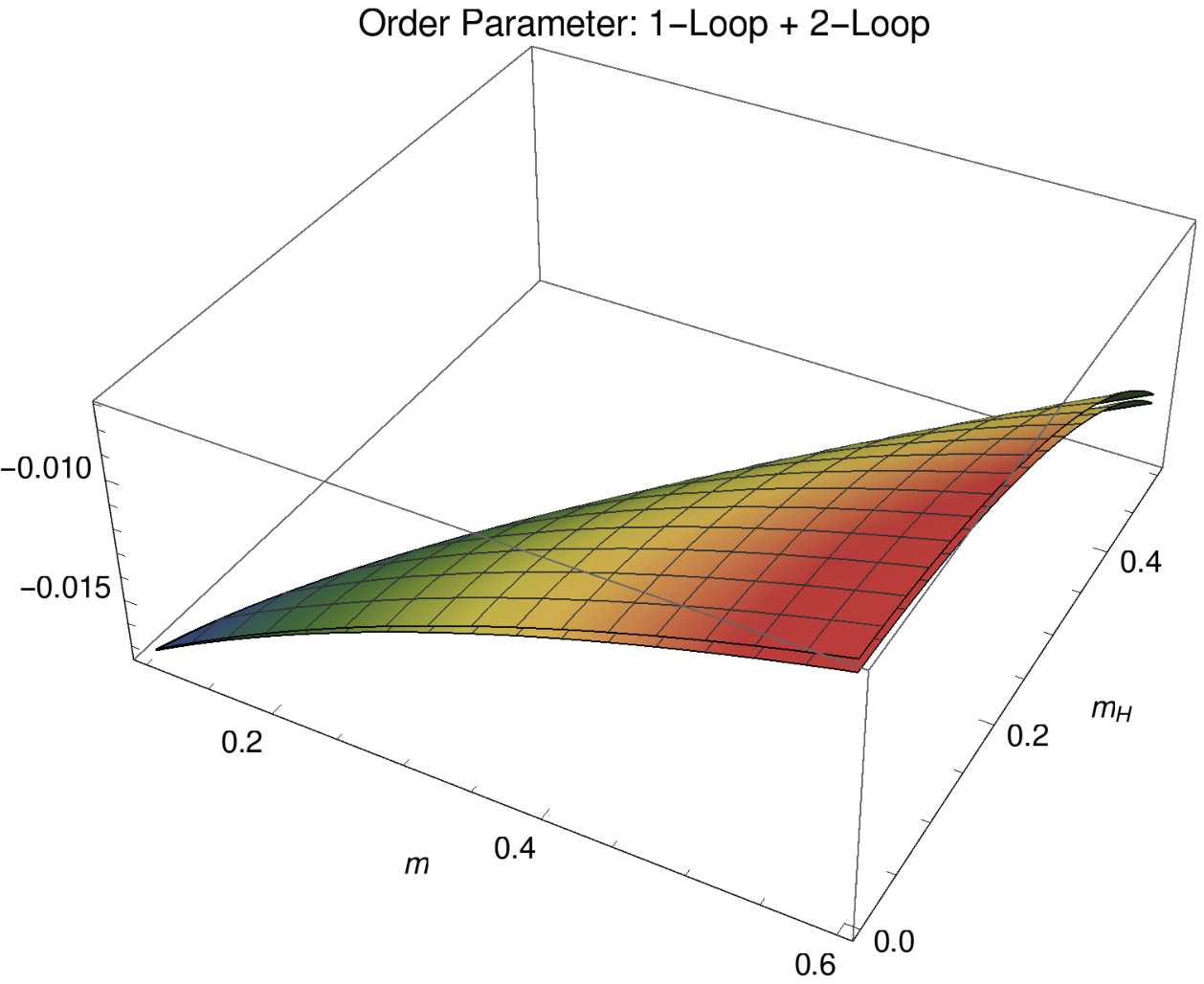}}
\end{center}
\caption{[Color online] Staggered magnetization at the temperatures $t=0.3$ (left) and $t=0.5$ (right): Dependence of $\xi_{M_s}$ on the
magnetic ($m_H$) and staggered ($m$) field.}
\label{figure3}
\end{figure}

At finite temperature, the behavior of the order parameter in external magnetic and staggered fields is governed by the Bose gas
contribution, as Fig.~\ref{figure3} illustrates. The plots depict the dimensionless quantity
\begin{equation}
\xi_{M_s}(T,H_s,H) = \frac{1}{M_s} \, \Big\{ {\tilde \sigma}_1 T^2 + {\tilde \sigma}_2 T^4 \Big\}
\end{equation}
for the two temperatures $T/\sqrt{\rho_s}=0.3$ and $T/\sqrt{\rho_s}=0.5$. Maximal and minimal surfaces have been obtained by scanning
${\overline k}_1$ and ${\overline k}_2$ as described previously. Indeed, the surface $\xi_{M_s}(T,H_s,H)$ corresponding to the Bose gas term
(${\tilde \sigma}_1 T^2$) is only slightly modified by the NLO effective constants: the two-loop contribution (${\tilde \sigma}_2 T^4$) is
small. The fact that the quantity $\xi_{M_s}$ is negative throughout parameter space, goes along with intuition: when temperature is raised
-- while magnetic and staggered field strengths held constant -- the order parameter decreases because thermal fluctuations become
stronger. In weak staggered fields the effect is most pronounced. 

\section{Magnetization}
\label{mag}

The magnetization
\begin{equation}
M(T,H_s,H) = - \frac{\partial z(T,H_s,H)}{\partial H}
\end{equation}
follows the low-temperature series
\begin{equation}
\label{magnetizationAF}
M(T,H_s,H) = M(0,H_s,H) + {\hat \sigma}_1 T^2 + {\hat \sigma}_2 T^4 + {\cal O}(T^6) \, ,
\end{equation}
where the respective coefficients are
\begin{eqnarray}
{\hat \sigma}_1(T,H_s,H) & = & \sqrt{\rho_s} \, t^2 \frac{\partial {\hat h}_0}{\partial m_H} \, , \nonumber \\
{\hat \sigma}_2(T,H_s,H) & = & \frac{1}{\sqrt{\rho_s}} \, \Bigg\{ - t^2 {\hat h}_1 \frac{\partial {\hat h}_0}{\partial m_H}
- m_H t^2 \frac{\partial {\hat h}_1}{\partial m_H} \frac{\partial {\hat h}_0}{\partial m_H}
- m_H t^2 {\hat h}_1 \frac{\partial^2 {\hat h}_0}{\partial m^2_H}
+ 2 m_H {({\hat h}_1)}^2 \nonumber \\
& & + 2 m^2_H {\hat h}_1 \frac{\partial {\hat h}_1}{\partial m_H}
+ \frac{{\overline k}_2-{\overline k}_1}{16 \pi^2} \, \frac{m^4}{t^2} \, \frac{\partial {\hat h}_1}{\partial m_H}
+ \frac{{\overline k}_1}{16 \pi^2} \, m^2 \, \frac{\partial {\hat h}_0}{\partial m_H} \nonumber \\
& & + \frac{{\overline k}_1}{16 \pi^2} \, m^2 m_H \, \frac{\partial^2 {\hat h}_0}{\partial m^2_H}
- \frac{{\overline k}_1}{4 \pi^2} \, \frac{m^2 m_H}{t^2} \, {\hat h}_1
- \frac{{\overline k}_1}{8 \pi^2} \, \frac{m^2 m^2_H}{t^2} \, \frac{\partial {\hat h}_1}{\partial m_H} \Bigg\} \, .
\end{eqnarray}
The order-$T^2$ term refers to the free magnon gas, while the $T^4$-term contains the spin-wave interaction. Regarding the zero-temperature
magnetization $M(0,H_s,H)$, according to Eq.~(\ref{z0}), the magnetic field only shows up at NNLO in the free energy density $z_0$, such
that the magnetization at $T$=0 is
\begin{equation}
M(0,H_s,H) = {\cal O}(p^6) \, ,
\end{equation}
i.e., negligible, and -- due to the unknown numerical values of the NNLO effective constants in ${\cal L}^6_{eff}$ -- also beyond the scope
of the present analysis.

\begin{figure}
\begin{center}
\hbox{
\includegraphics[width=8.2cm]{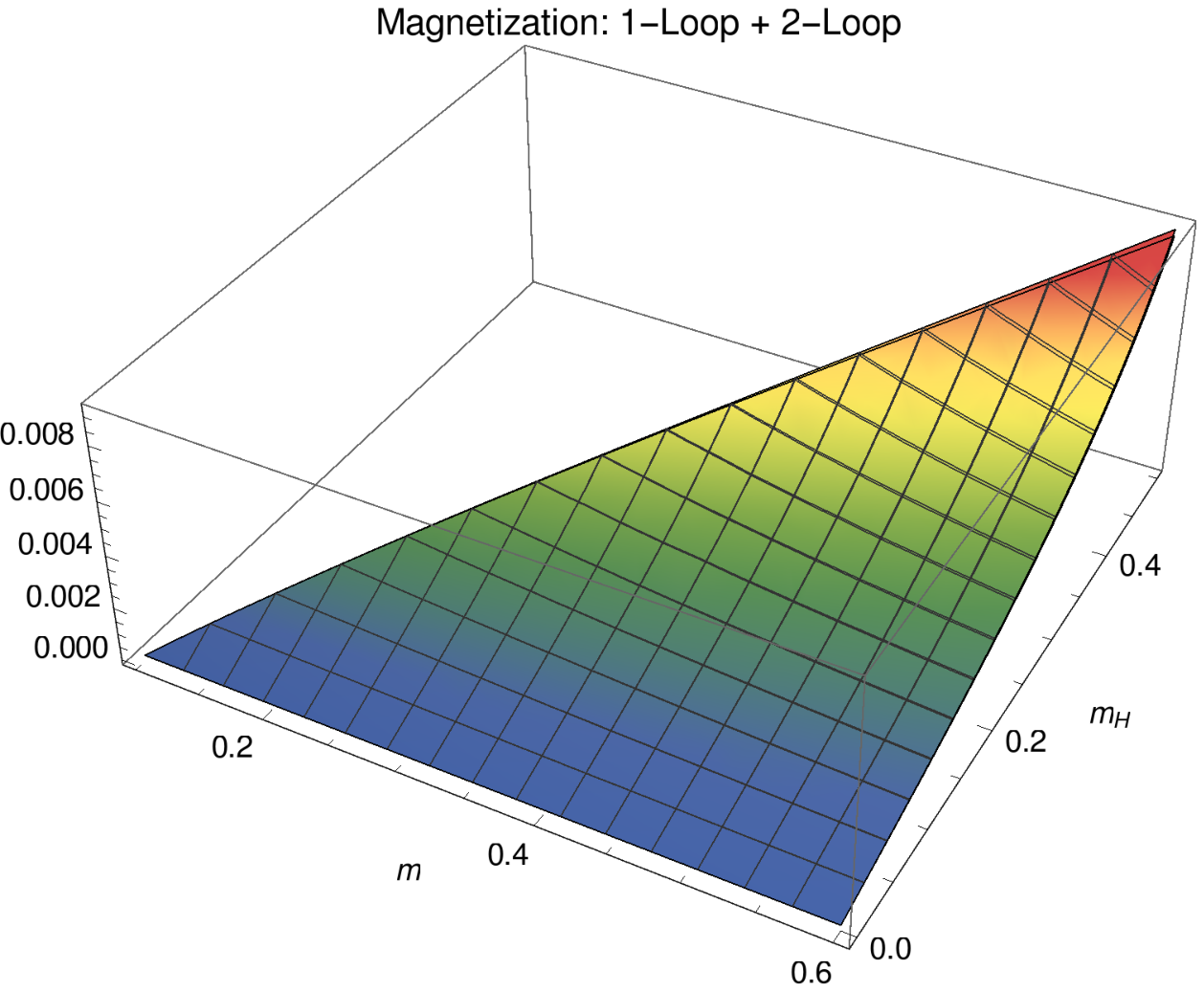} 
\includegraphics[width=8.2cm]{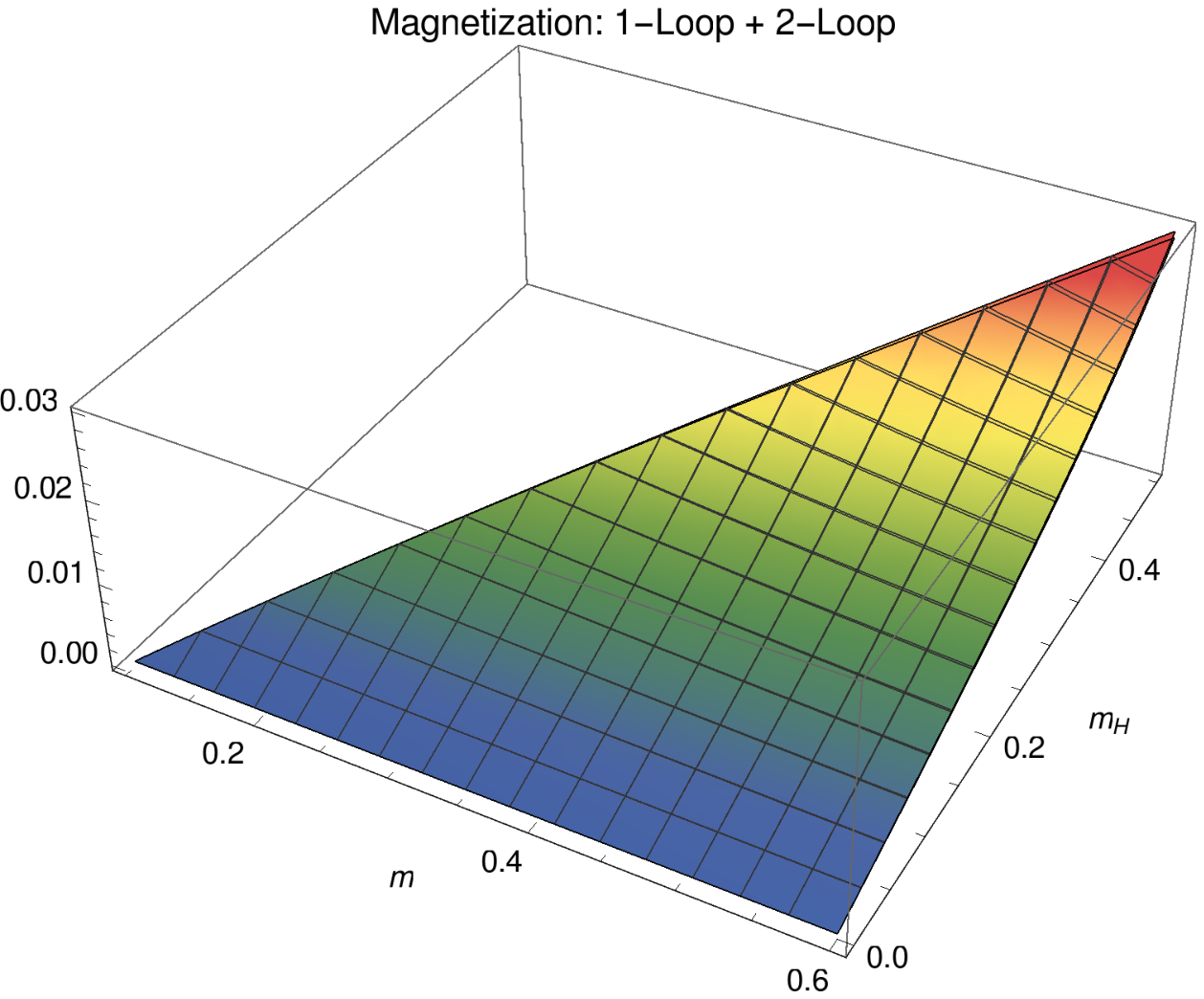}}
\end{center}
\caption{[Color online] Magnetization at the temperatures $t=0.3$ (left) and $t=0.5$ (right): Dependence of $\xi_{M}$ on the magnetic
($m_H$) and staggered ($m$) field.}
\label{figure4}
\end{figure}

Let us consider the behavior at finite temperature. In Fig.~\ref{figure4}, we plot the normalized finite-temperature magnetization,
\begin{equation}
\xi_{M}(T,H_s,H) = \frac{{\hat \sigma}_1 T^2 + {\hat \sigma}_2 T^4}{{\rho_s}^{\frac{3}{2}}} \, ,
\end{equation}
for the two temperatures $T/\sqrt{\rho_s}=0.3$ and $T/\sqrt{\rho_s}=0.5$. One first observes that the magnetization remains zero when only
the staggered field is present. In order to induce a net magnetization, an asymmetric situation must be generated. The staggered field
though affects sublattice $A$ and sublattice $B$ of our bipartite system in a symmetric manner: it points into the same direction as the
up-spins on sublattice $A$, and it also points into the same direction as the down-spins on sublattice $B$. As such, quantum fluctuations
on sublattice $A$ and $B$ are suppressed the very same way by the staggered field -- as a result, no net magnetization can emerge.

The effect of the uniform magnetic field is qualitatively different: the net field that experience up-spins is $H_s+H$, while down-spins
sense the weaker field $H_s-H$. Quantum fluctuations of the up-spins are then more suppressed than quantum fluctuations of the down spins
-- as a consequence a net magnetization in the direction of the magnetic field is generated. According to Fig.~\ref{figure4}, the induced
magnetization, measured by the quantity $\xi_{M}(T,H_s,H)$, is small. One further notices that the behavior of the magnetization at finite
temperature is essentially described by the one-loop contribution. The minimal and maximal surfaces obtained by scanning ${\overline k}_1$
and ${\overline k}_2$ are barely visible, i.e., they almost coincide with the surface for ${\hat \sigma}_1 T^2/{\rho_s^{3/2}}$.

Astonishing, however, is the fact that the quantity $\xi_{M}(T,H_s,H)$ is larger at $t=0.5$ than at $t=0.3$ for any given point $m_H,m$ in
parameter space. This means that -- magnetic and staggered fields held fixed -- the net magnetization is larger at higher temperatures. One
would rather expect the net magnetization to decrease as thermal fluctuations become more prominent at higher temperatures. This somehow
counterintuitive observation is implicitly contained in Eq.~(7.4.126) of Ref.~\citep{Nol86}, although the phenomenon has not been pointed
out there.\footnote{W.\ Nolting, private communication.}

\section{Conclusions}
\label{conclusions}

Within the framework of magnon effective field theory, we have derived the partition function for antiferromagnetic solids in mutually
parallel magnetic and staggered fields up to two-loop order where the magnon-magnon interaction becomes relevant. As it turned out, the
behavior of the system is dominated by the one-loop contribution, i.e., interaction effects in the thermodynamic quantities are small. To
establish this result, we had to estimate and scan the numerical values of next-to-leading order renormalized effective constants.

For fixed magnetic and staggered fields, the finite-temperature order parameter drops when temperature increases. The decrease is larger in
weak staggered fields due to the fact that thermal fluctuations win over the suppression of quantum fluctuations caused by the staggered
field. Interestingly, for fixed magnetic and staggered fields, the finite-temperature magnetization grows when temperature increases. This
phenomenon that appears to be rather counterintuitive, emerges at the one-loop level, i.e., it is not induced by spin-wave interactions,
but refers to the free magnon gas.

To have a more quantitative picture of the magnitude of the two-loop corrections -- going beyond the scans -- the actual numerical values
of next-to-leading order effective constants should be known. One way is to numerically simulate -- on the basis of efficient loop-cluster
algorithms -- a specific system such as the simple cubic antiferromagnet, and then extract the NLO effective constants to make the
effective field theory results even more predictive. Work in this direction is currently in progress.

\section*{Acknowledgments}
The author thanks W.\ Nolting for correspondence.

\begin{appendix}

\section{Explicit evaluation of the free energy density}
\label{appendixA}

In this appendix we complement the results presented in the main body of the paper by explicitly deriving the individual contributions to
the free energy density.

\subsection{Tree graphs 2, 4B, and 6C}

The leading temperature-independent contribution in the free energy density, originating from graph 2, is finite and reads\footnote{Recall
that the magnon mass squared is proportional to the staggered field: $M^2=M_s H_s / \rho_s$.}
\begin{equation}
z_2 = - \rho_s M^2 \, .
\end{equation}

The tree-graph 4B involves the next-to-leading order Lagrangian ${\cal L}^4_{eff}$ and gives rise to
\begin{equation}
z_{4B} = - (k_2 + k_3) M^4 \, .
\end{equation}
This expression is singular on account of the NLO effective constants $k_2$ and $k_3$, and hence needs to be renormalized. Following the
standard convention (see Ref.~\cite{Hof17b}), the NLO effective constants are written as
\begin{equation}
k_2 = {\tilde \gamma}_4 \Big( \lambda + \frac{{\overline k}_2}{32 \pi^2} \Big) \, , \quad
k_3 = {\tilde \gamma}_5 \Big( \lambda + \frac{{\overline k}_3}{32 \pi^2} \Big) \, ,
\end{equation}
and one ends up with
\begin{equation}
z_{4B} = - \frac{M^4}{32 \pi^2} \Big( {\tilde \gamma}_4 {\overline k}_2 + {\tilde \gamma}_5 {\overline k}_3 \Big)
- ({\tilde \gamma}_4 + {\tilde \gamma}_5) M^4 \lambda \, .
\end{equation}
Whereas the renormalized effective constants ${\overline k}_2$ and ${\overline k}_3$ -- as well as the dimensionless coefficients
${\tilde \gamma}_4$ and ${\tilde \gamma}_5$ -- are finite, the parameter $\lambda$,
\begin{eqnarray}
\label{lambda}
\lambda & = & \mbox{$ \frac{1}{2}$} \, (4 \pi)^{-d/2} \, \Gamma(1-{\mbox{$ \frac{1}{2}$}}d) M^{d-4} \nonumber\\
& = & \frac{M^{d-4}}{16{\pi}^2} \, \Bigg[ \frac{1}{d-4} - \mbox{$ \frac{1}{2}$} \{ \ln{4{\pi}} + {\Gamma}'(1) + 1 \}
+ {\cal O}(d-4) \Bigg] \, ,
\end{eqnarray}
becomes singular in the physical limit $d \to 4$. As we show below, the same singularity emerges in the one-loop graph 4A, such that in the
sum of order-$p^4$ graphs -- $z_{4A} + z_{4B}$ -- the divergences cancel. Using the relation
\begin{equation}
{\tilde \gamma}_4 + {\tilde \gamma}_5 = 1 \, ,
\end{equation}
we obtain
\begin{equation}
z_{4B} = - \frac{M^4}{32 \pi^2} \Big( {\tilde \gamma}_4 ( {\overline k}_2 - {\overline k}_3) + {\overline k}_3 \Big) - M^4 \lambda \, .
\end{equation}

Finally the order-$p^6$ tree graph 6C involves the next-to-next-to-leading order effective Lagrangian ${\cal L}^6_{eff}$, generating terms
proportional to $M^6, M^4 H^2, M^2 H^4$ and $H^6$ in the free-energy density contribution $z_{6C}$, along with additional NNLO effective
constants (contained in ${\cal L}^6_{eff}$) whose numerical values are tiny and a priori unknown. Since we are furthermore dealing with a
temperature-independent contribution to the free energy density, it is not illuminating to provide an explicit expression for $z_{6C}$.

\subsection{One-loop graph 4A}

Here we have to evaluate the functional integral $J$
\begin{equation}
J = \int [ \mbox{d} U ] \exp \! \Big[ - \int \! \mbox{d}^d x \, {\cal L}_{kin} \Big] \, , 
\end{equation}
in order to obtain the free energy density $z_{4A}$ as
\begin{equation}
z_{4A} = - \frac{1}{V_d} \, \log J \, .
\end{equation}
The quantity $V_d$ represents the Euclidean volume. Rather than directly evaluating $J$, is convenient to consider the derivative,
\begin{eqnarray}
\frac{\partial}{\partial M^2} \, J & = & - \int [ \mbox{d} u ] [ \mbox{d} u^{*} ] \exp \! \Big[ - \int \! \mbox{d}^d x
\, {\cal L}_{kin} \Big] \, \frac{\rho_s}{2} \! \int \! \mbox{d}^d x \, u u^{*} \nonumber \\
& & = - V_d J \, {\hat G}(0) \, ,
\end{eqnarray}
with the kinetic magnon contribution from the leading-order effective Lagrangian ${\cal L}^2_{eff}$,
\begin{equation}
{\cal L}_{kin} = \mbox{$ \frac{1}{2}$} \rho_s \partial_{\mu} u \partial_{\mu} u^{*} + \mbox{$ \frac{1}{2}$} \rho_s M^2 u u^{*}
- \mbox{$ \frac{1}{2}$} \rho_s H (u^{*} \partial_0 u - u \partial_0 u^{*}) - \mbox{$ \frac{1}{2}$} \rho_s H^2 u u^{*} \, .
\end{equation}
The free energy density is then obtained by integrating over $M^2$,
\begin{equation}
z_{4A} = - {\hat G}(0) \, .
\end{equation}
Recall that the thermal propagators $G^{\pm}(x)$,
\begin{equation}
G^{\pm}(x) = \sum_{n = - \infty}^{\infty} \Delta^{\pm}({\vec x}, x_4 + n \beta) \, , \qquad \beta = \frac{1}{T} \, ,
\end{equation}
can be decomposed into a zero-temperature piece and a finite-temperature piece. At the origin $x$=0, the thermal propagators coincide,
\begin{equation}
G^{+}(0) = G^{-}(0) \equiv {\hat G}(0) \, ,
\end{equation}
and the decomposition is
\begin{equation}
{\hat G}(0) = {\hat g}_1 + \Delta(0) \, .
\end{equation}
The explicit expressions for the Bose function ${\hat g}_1$ and the zero-temperature propagator $\Delta(0)$ are provided in
Eq.~(\ref{g1BoseD3}) and Eq.~(\ref{regprop}), respectively. The one-loop free energy density $z_{4A}$ then amounts to
\begin{eqnarray}
z_{4A} & = & - {\hat g}_0 - \frac{M^d}{2^d \pi^{d/2}} \, \Gamma\Big(-\frac{d}{2}\Big) \, , \nonumber \\
& = & - {\hat g}_0 + \frac{4}{d} M^4 \lambda \, ,
\end{eqnarray}
where ${\hat g}_0$ is defined in Eq.~(\ref{g0Bose}).

\subsection{One-loop graph 6B}

The next-to-leading order Lagrangian ${\cal L}^4_{eff}$ gives rise to the following terms quadratic in the magnon fields,
\begin{equation}
{\cal L}_{6B} =  k_1 M^2 \partial_{\mu} u \partial_{\mu} u^{*} + k_2 M^4 u u^{*} - k_1 M^2 H^2 u u^{*}
+ k_1  M^2 H ( u \partial_0 u^{*} - \partial_0 u u^{*} ) \, .
\end{equation}
Closing the magnon loop, i.e., contracting the magnon fields, the free energy density $z_{6B}$ amounts to\footnote{Note that time
derivatives refer to Euclidean time $x_4 = i t$.}
\begin{equation}
z_{6B} =  (k_2 - k_1) \frac{M^4}{\rho_s} \Big( G^{+}(0) + G^{-}(0) \Big) + 2 k_1 \frac{M^2 H}{\rho_s}
{\Big( {\dot G}^{+}(x) - {\dot G}^{-}(x) \Big)}_{|x=0} \, .
\end{equation}
Following the standard convention (see Ref.~\cite{Hof17b}), the NLO effective constant $k_1$ is expressed as
\begin{equation}
k_1 = {\tilde \gamma}_3 \Big( \lambda + \frac{{\overline k}_1}{32 \pi^2} \Big) \, .
\end{equation}
Using the relations
\begin{eqnarray}
\label{propRelations}
{\Big\{G^+(x) + G^-(x) \Big\}}_{|x=0} & = & 2 {\hat g}_1 + 4 M^2 \lambda \, , \\
{\Big\{{\dot G}^+(x) - {\dot G}^-(x) \Big\}}_{|x=0} & = & \frac{\partial}{\partial H} \, {\hat g}_0
- 2 H {\hat g}_1 - 4 M^2 H \lambda \, , \nonumber
\end{eqnarray}
as well as the fact that the numerical coefficients ${\tilde \gamma}_3$ and $ {\tilde \gamma}_4$ are identical (see Ref.~\citep{Hof17b}),
\begin{equation}
{\tilde \gamma}_3 - {\tilde \gamma}_4 = 0 \, ,
\end{equation}
we end up with
\begin{eqnarray}
z_{6B} & = & {\tilde \gamma}_3 ( {\overline k}_2 - {\overline k}_1) \frac{M^4}{16 \pi^2 \rho_s} \, {\hat g}_1
+ {\tilde \gamma}_3 ( {\overline k}_2 - {\overline k}_1) \frac{M^6}{8 \pi^2 \rho_s} \lambda \nonumber \\
& & + {\tilde \gamma}_3  {\overline k}_1 \frac{M^2 H}{16 \pi^2 \rho_s} \, \Big( \frac{\partial {\hat g}_0}{\partial H} - 2 H {\hat g}_1 \Big)
- {\tilde \gamma}_3 {\overline k}_1 \frac{M^4 H^2}{4 \pi^2 \rho_s} \lambda \nonumber \\
& & + 2 {\tilde \gamma}_3 \frac{M^2 H}{\rho_s} \, \Big( \frac{\partial {\hat g}_0}{\partial H} - 2 H {\hat g}_1 \Big) \lambda
- 8 {\tilde \gamma}_3 \frac{M^4 H^2}{\rho_s} \, \lambda^2 \, .
\end{eqnarray}
This expression is singular due to the parameter $\lambda$. The point is that analogous infinities show up in the two-loop graph 6A such
that the order-$p^6$ singularities in the sum $z_{6A} + z_{6B} + z_{6C}$ cancel alltogether (see below). 

\subsection{Two-loop graph 6A}

The leading-order effective Lagrangian ${\cal L}^2_{eff}$ generates the following terms quartic in the magnon fields,
\begin{equation}
{\cal L}_{6A} = \mbox{$ \frac{1}{4}$} \rho_s \partial_{\mu} u \partial_{\mu} u^{*} u u^{*}
+ \mbox{$ \frac{1}{8}$} \rho_s \partial_{\mu} u^{*} u \partial_{\mu} u^{*} u
+ \mbox{$ \frac{1}{8}$} \rho_s \partial_{\mu} u u^{*} \partial_{\mu} u u^{*}
+ \mbox{$ \frac{1}{8}$} \rho_s M^2 u u^{*} u u^{*} \, .
\end{equation}
We then evaluate the functional integral
\begin{equation}
J_{6A} = \int [ \mbox{d} u ]  [ \mbox{d} u^{*} ] \exp \! \Big[ - \int \! \mbox{d}^d x \, {\cal L}_{kin} \Big] \, \! \int \! \mbox{d}^d x \,
{\cal L}_{6A} \, ,
\end{equation}
and obtain the two-loop contribution to the free energy density as
\begin{eqnarray}
\label{z6AAppendix}
z_{6A} & = & \frac{H}{2 \rho_s} {\Big( {\dot G}^+(x) - {\dot G}^-(x) \Big)}_{|x=0} \Big( G^+(0) + G^-(0) \Big)
+ \frac{H^2}{4 \rho_s} {\Big( G^+(0) + G^-(0)  \Big)}^2 \nonumber \\
& = & \frac{H}{\rho_s} \, {\hat g}_1 \, \frac{\partial {\hat g}_0}{\partial H}
- \frac{H^2}{\rho_s}{( {\hat g}_1)}^2
- \frac{4 H^2 M^2}{\rho_s} \, {\hat g}_1 \, \lambda
+ \frac{2 H M^2}{\rho_s} \, \frac{\partial {\hat g}_0}{\partial H} \, \lambda
- \frac{4 H^2 M^4}{\rho_s} \, \lambda^2 \, .
\end{eqnarray}
In the above calculation we have used the propagator equations
\begin{equation}
\Big\{ \Box - M^2 \pm 2 H \partial_{x_4} + H^2 \Big\} \, G^{\pm}(x)_{|_{x=0}} = 0 \, ,
\end{equation}
as well as the relations (\ref{propRelations}).

\subsection{Total two-loop free energy density}

Collecting all contributions up to two-loop order, we get
\begin{eqnarray}
\label{z2loopSum}
z & = & z_2 + z_{4A} + z_{4B} +z_{6A} + z_{6B} + z_{6C} \nonumber \\
& = & - \rho_s M^2 - {\hat g}_0 - \frac{M^4}{64 \pi^2}
- \frac{M^4}{32 \pi^2} \Big( {\tilde \gamma}_4 ( {\overline k}_2 - {\overline k}_3) + {\overline k}_3 \Big) \nonumber \\
& & + \frac{H}{\rho_s} \, {\hat g}_1 \, \frac{\partial {\hat g}_0}{\partial H}
- \frac{H^2}{\rho_s}{( {\hat g}_1)}^2
- \underline{\frac{4 H^2 M^2}{\rho_s} \, {\hat g}_1 \, \lambda}
+ \underline{\frac{2 H M^2}{\rho_s} \, \frac{\partial {\hat g}_0}{\partial H} \, \lambda}
- \frac{4 H^2 M^4}{\rho_s} \, \lambda^2 \nonumber \\
& & + {\tilde \gamma}_3 ( {\overline k}_2 - {\overline k}_1) \frac{M^4}{16 \pi^2 \rho_s} \, {\hat g}_1
+ {\tilde \gamma}_3 ( {\overline k}_2 - {\overline k}_1) \frac{M^6}{8 \pi^2 \rho_s} \lambda
+ {\tilde \gamma}_3  {\overline k}_1 \frac{M^2 H}{16 \pi^2 \rho_s} \, \Big( \frac{\partial {\hat g}_0}{\partial H} - 2 H {\hat g}_1 \Big)
\nonumber \\
& & - {\tilde \gamma}_3 {\overline k}_1 \frac{M^4 H^2}{4 \pi^2 \rho_s} \lambda
+ \underline{2 {\tilde \gamma}_3 \frac{M^2 H}{\rho_s} \, \Big( \frac{\partial {\hat g}_0}{\partial H} - 2 H {\hat g}_1 \Big) \lambda}
- 8 {\tilde \gamma}_3 \frac{M^4 H^2}{\rho_s} \lambda^2 + z_{6C} \, .
\end{eqnarray}
While the general constraints
\begin{equation}
{\tilde \gamma}_3 = {\tilde \gamma}_4 \, , \qquad {\tilde \gamma}_4 + {\tilde \gamma}_5 = 1 \, ,
\end{equation}
have to be satisfied (see Ref.~\cite{Hof17b}), we now choose the specific value ${\tilde \gamma}_3 = -1$. In this case, the underlined
divergent and temperature-dependent terms in Eq.~(\ref{z2loopSum}) mutually cancel. With the fixed set of coefficients,
\begin{equation}
{\tilde \gamma}_3 = -1 \, , \qquad {\tilde \gamma}_4 = -1 \, , \qquad  {\tilde \gamma}_5 = 2 \, ,
\end{equation}
the final representation for the two-loop free energy density, decomposed into the temperature-independent piece $z_0$, and the
finite-temperature piece $z^T$,
\begin{equation}
z = z_0 + z^T \, ,
\end{equation}
takes the form
\begin{eqnarray}
\label{TwoLoopz0zT}
z_0 & = & - M_s H_s + \frac{ {\overline k}_2 - 2{\overline k}_3}{32 \pi^2} \, \frac{M^2_s H^2_s}{\rho^2_s}
- \frac{M^2_s H^2_s}{64 \pi^2 \rho^2_s} + {\cal O}(p^6) \, , \nonumber \\
z^T & = & - {\hat g}_0 + \frac{H}{\rho_s} \, {\hat g}_1 \, \frac{\partial {\hat g}_0}{\partial H}
- \frac{H^2}{\rho_s}{( {\hat g}_1)}^2
- \frac{{\overline k}_2 - {\overline k}_1}{16 \pi^2} \frac{M^2_s H^2_s}{\rho^3_s} \, {\hat g}_1
- \frac{{\overline k}_1}{16 \pi^2} \, \frac{H M_s H_s}{\rho^2_s} \frac{\partial {\hat g}_0}{\partial H} \nonumber \\
& & + \frac{{\overline k}_1}{8 \pi^2} \, \frac{H^2 M_s H_s}{\rho^2_s} \, {\hat g}_1 + {\cal O}(p^8) \, .
\end{eqnarray}
Here in the final result we have re-expressed the magnon mass through the staggered field as $M^2=M_s H_s/\rho_s$.
  
\end{appendix}

\end{document}